\documentclass[%
reprint,
nobibnotes,
amsmath,amssymb,
prb,
floatfix,
]{revtex4-2}

\usepackage{graphicx}
\usepackage{dcolumn}
\usepackage{bm}
\usepackage{mathtools}
\usepackage{braket}
\usepackage[caption=false, position=top]{subfig}
\usepackage{lipsum}
\usepackage{hyperref}
\usepackage{booktabs}
\usepackage{array}
\usepackage{multirow}
\usepackage{longtable}
\makeatletter
\def\nobreakhline{%
  \noalign{\ifnum0=`}\fi
    \penalty\@M
    \futurelet\@let@token\LT@@nobreakhline}
\def\LT@@nobreakhline{%
  \ifx\@let@token\hline
    \global\let\@gtempa\@gobble
    \gdef\LT@sep{\penalty\@M\vskip\doublerulesep}
  \else
    \global\let\@gtempa\@empty
    \gdef\LT@sep{\penalty\@M\vskip-\arrayrulewidth}
  \fi
  \ifnum0=`{\fi}%
  \multispan\LT@cols
     \unskip\leaders\hrule\@height\arrayrulewidth\hfill\cr
  \noalign{\LT@sep}%
  \multispan\LT@cols
     \unskip\leaders\hrule\@height\arrayrulewidth\hfill\cr
  \noalign{\penalty\@M}%
  \@gtempa}
\makeatother

\newcolumntype{P}[1]{>{\centering\arraybackslash}p{#1}}
\newcolumntype{M}[1]{>{\centering\arraybackslash}m{#1}}

\renewcommand{\bar}[1]{\overline{#1}}

\renewcommand{\vec}[1]{\boldsymbol{\mathrm{#1}}}

\newcommand{\abs}[1]{|#1|}

\newcommand{\norm}[1]{||{#1}||}
\newcommand{\Norm}[1]{\left|\left|{#1}\right|\right|}

\begin{document}

\preprint{APS/123-QED}

\title{Dispersion-corrected Machine Learning Potentials for 2D van der Waals Materials}

\author{Mikkel Ohm Sauer$^{1}$}
\author{Peder Meisner Lyngby$^{1}$}
\author{Kristian Sommer Thygesen$^{1}$}
\email{thygesen@fysik.dtu.dk}
\affiliation{$^1$CAMD, Department of
Physics, Technical University of Denmark, DK - 2800 Kongens Lyngby,
Denmark}

\date{\today}

\begin{abstract}
Machine-learned interatomic potentials (MLIPs) based on message passing neural networks hold promise to enable large-scale atomistic simulations of complex materials with \emph{ab initio} accuracy. A number of MLIPs trained on energies and forces from density functional theory (DFT) calculations employing semi-local exchange-correlation (xc) functionals have recently been introduced. Here, we benchmark the performance of six dispersion-corrected MLIPs on a dataset of van der Waals heterobilayers containing between 4 and 300 atoms in the moiré cell. Using various structure similarity metrics, we compare the relaxed heterostructures to the ground truth DFT results. With some notable exceptions, the model precisions are comparable to the uncertainty on the DFT results stemming from the choice of xc-functional. We further explore how the structural inaccuracies propagate to the electronic properties, and find excellent performance with average errors on band energies as low as 35 meV. Our results demonstrate that recent MLIPs after dispersion corrections are on par with DFT for general vdW heterostructures, and thus justify their application to complex and experimentally relevant 2D materials. 
\end{abstract}

\maketitle

\section{Introduction}
\begin{figure*}
    \centering
    \includegraphics[width=\linewidth]{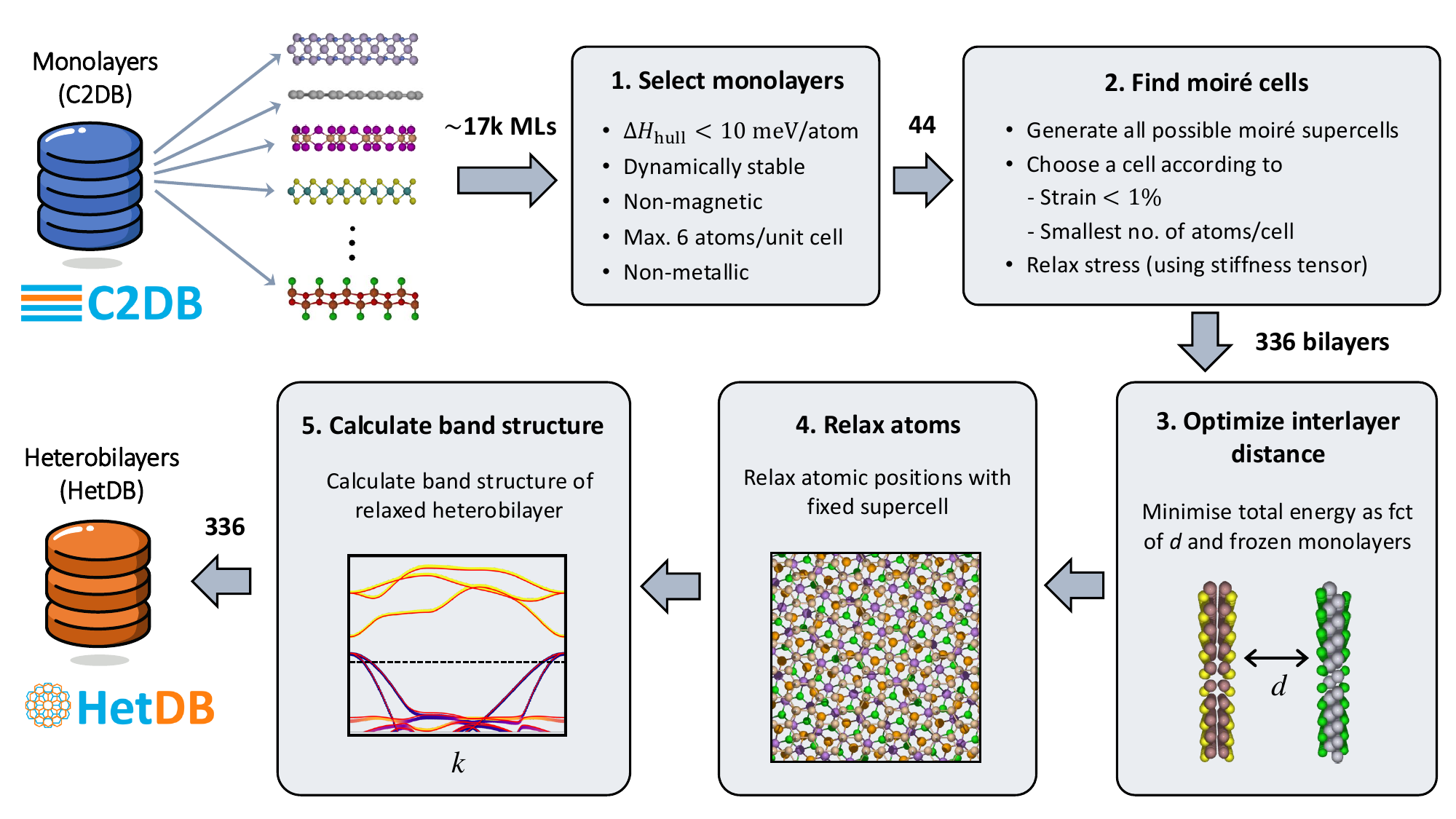}
    \caption{The workflow used to construct the vdW heterobilayers and calculate their electronic band structure. As a first step 44 monolayers are extracted from C2DB all fulfilling the requirements in box 1. Next, moiré supercell basis vectors for all possible combinations of the 44 monolayers that satisfy the conditions in box 2 are determined. This results in a total of 336 heterobilayers. For each of these bilayers the optimal interlayer distance  is determined as the minimum of the binding energy as the monolayers are displaced rigidly relative to each other. Next, the atomic positions are relaxed within a fixed supercell. Finally, the electronic band structures of all optimized heterobilayers is calculated using DFT-PBE with a double-zeta polarized LCAO basis set. The atomic structures of the relaxed heterobilayers and their electronic band structures are available in the online database HetDB.}
    \label{fig:intro}
\end{figure*}

Atomically thin two-dimensional (2D) crystals, like graphene and transition metal dichalcogenides (TMDs), have emerged as a promising class of materials with unique physical properties\cite{geim2007rise,mak2010atomically,ferrari2015science}. Many of these properties can be enhanced, and new ones emerge, when the individual 2D layers are combined into van der Waals (vdW) heterostructures\cite{geim2013van}. Even the seemingly simple class of naturally stacked homobilayers\cite{pakdel2024high} presents novel features, such as sliding ferroelectricity\cite{yasuda2021stacking,rogee2022ferroelectricity} and electrically tunable interlayer excitons\cite{peimyoo2021electrical}. In twisted homobilayers or lattice mismatched heterobilayers, the periodic moiré potential can influence exciton physics\cite{alexeev2019resonantly} and introduce flat bands leading to unconventional superconductivity\cite{cao2018unconventional}, Mott insulators\cite{po2018origin}, and novel types of magnetism\cite{cheng2023electrically}.

Identifying the optimal vdW heterostructure for a specific application or physical phenomenon is in general a tremendous challenge due to the huge size of the 2D materials space. Around 250 2D materials have so far been experimentally produced in mono- or few-layer form\cite{amin}, while computational databases hold the structures of thousands of stable and potentially synthesizable monolayer crystals\cite{article:C2DB,lyngby2024ab,zhou20192dmatpedia}. It is expected that most of these monolayers can be combined in various stacking configurations yielding millions of possible heterostructures.    

Besides the combinatorial challenge, a major obstacle currently hindering an efficient computational exploration of the vdW heterostructure space, is the large number of atoms contained in a single unit cell (from hereon referred to as the moiré cell). Indeed, they easily contain hundreds or thousands of atoms rendering the use of \emph{ab initio} methods such as density functional theory (DFT) a daunting challenge. The recently introduced machine learning interatomic potentials (MLIPs) based on message passing neural networks could help to overcome this barrier by enabling efficient optimization of the atomic structure. Since actual applications depend on the physical properties of the material rather than its detailed atomic structure, the relevant question in this context is how the inaccuracies in the optimized geometries are reflected in the electronic structure of the materials. 

In this work, we benchmark the performance of six different MLIPs for describing general vdW heterostructures. To describe the crucial non-local vdW interaction acting between the layers, we augment the MLIPs by the D3 dispersion correction of Grimme \emph{et al.}\cite{grimme2010consistent}. 
We study 336 heterobilayers comprising a total of 44 different monolayers. The bilayers are selected such that a moiré cell with less than 300 atoms and in-plane strain below 1\% can be constructed. The intermediate size of the benchmark structures makes it possible to assess the accuracy of the MLIPs by comparing to well converged PBE-D3 calculations representing the ground truth for the models. The structural similarities are evaluated using a number of metrics covering the interlayer distance and the internal atomic positions and stress of the monolayers. Finally, we calculate the electronic band structures of the heterobilayers relaxed with MLIP-D3 and DFT-D3, respectively, and evaluate the difference employing both quantitative and qualitative measures. All the relaxed bilayers and their calculated band structures are available in the \emph{HetDB} database\cite{hetdb}, which is integrated with the C2DB and other 2D materials databases. 
This workflow is summarized in Fig. \ref{fig:intro}, and further details about the individual parts can be found in the Methods section \ref{sec:methods}.

\section{Definition of metrics}
In order to compare and benchmark the different dispersion corrected MLIPs, we use a set of \emph{structural} metrics defined to capture the distance between the two layers, the reorganization of the internal atomic structure of the monolayers, and the amount of in-plane stress. Additionally, we define two \emph{electronic} metrics to describe the change in the electronic band structure.

As a measure of the interlayer distance, we use the mean $z$-distance between the atoms of the two layers. Specifically, we first define the average $z$-position of atoms in layer $i$,
\begin{equation}
\langle z_i \rangle = \frac{1}{N_i} \sum_{a\in \mathrm{cell}} z_{a,i},
\end{equation}
where the sum is over all atoms in layer $i$ and $N_i$ is the number of atoms of that layer in the moiré cell. The (average) interlayer distance for a given relaxation method $M$ is then
\begin{equation}
   d^{(M)} = \langle z_2 \rangle^{(M)} - \langle z_1\rangle^{(M)},
   \label{eq:avg_distance}
\end{equation}
and we define the \emph{interlayer distance metric} as
\begin{align}
    \Delta d_{\mathrm{inter}}^{(M)} &= d^{(M)} - d^{(\mathrm{PW})}, \label{eq:interlayer_delta} 
\end{align}
where 'PW' refers to the PBE-D3(PW) ground truth method.

To measure the variation in the atomic structure within the monolayers we first normalize the atomic positions of layer $i$ with respect to the average $z$-position of the layer
\begin{equation}
    \tilde{\vec{r}}_{a,i}=\vec{r}_{a,i}  - \langle z_i\rangle \vec{e}_z.
\end{equation}
We then define an \emph{intralayer metric} for method $M$ as the root-mean-square deviation (RMSD) of the atomic positions within the layers relative to the ground truth PBE-D3(PW) method,  
\begin{equation}\label{eq:intra}
    \Delta R_{\mathrm{intra}}^{(M)} = \sqrt{\frac{1}{N_1 + N_2}\sum_{i=1,2}\sum_a{\Norm{\tilde{\vec{r}}_{a,i}^{(M)} - \tilde{\vec{r}}_{a,i}^{(\mathrm{PW})}}^2}}.
\end{equation}

Next, in order to measure the amount of in-plane stress in the structures we use the \emph{stress metric} as
\begin{equation}\label{eq:stress_metric}
\Delta E_{\mathrm{stress}}^{(M)} = |E_{\mathrm{stress}}^{(M)}-E_{\mathrm{stress}}^{(\mathrm{PW})}|,
\end{equation}
where $E_{\mathrm{stress}}^{(M)}$ is the mechanical energy defined in Eq. (\ref{eq:mech}) (methods section \ref{sec:methods}) evaluated with method $M$ for the PBE-D3(PW) moiré cell.   


To complement the structural metrics introduced above, we introduce a set of electronic metrics. The \emph{band gap metric} is straightforwardly defined as,
\begin{align}\label{eq:gap_metric}
    \Delta E_{\text{gap}}^{(M)} &= E_{\text{gap}}^{(M)} - E_{\text{gap}}^{(\text{PW})}, 
\end{align}
where $E_{\mathrm{gap}}^{(M)}$ is the band gap calculated using one-shot PBE(LCAO) on the heterobilayer relaxed by method $M$. 

Finally, to compare the full band structures we use a \emph{band energy metric} defined as the RMSD of the individual band energies,
\begin{equation}\label{eq:band_metric}
    \Delta \varepsilon_{\mathrm{band}}^{(M)} = \sqrt{\frac{1}{20 N_k}\sum_k \sum_{n} (\varepsilon_{nk}^{(M)} - \varepsilon_{nk}^{(\mathrm{PW})})^2},
\end{equation}
where the sums run over $k$-points and the 10 highest valence bands and 10 lowest conduction bands, respectively. $\varepsilon_{nk}^{(M)}$ is a band energy calculated using one-shot PBE(LCAO) on the heterobilayer relaxed by method $M$.

\subsection{Machine learning models}
In this work, we have employed six MLIPs (including two different versions of the MACE potential). All the models have been trained on datasets containing PBE total energies and atomic forces of (mainly) inorganic bulk crystals. As such, by augmenting the model potentials with the D3 dispersion correction, they should yield PBE-D3 energies and forces. It should, however, be noted that these models have encountered a very limited amount of isolated 2D materials and vdW heterostructures during their training. All the models have been employed without any prior fine-tuning.  

\begin{itemize}
    \item Large MACE-MP-0 (L-MACE-MP) and Medium MACE-MPA-0 (M-MACE-MPA)\cite{batatia2024foundation}: MACE is designed around using an atomic cluster expansion as a local descriptor. The large MACE-MP-0 model utilizes message-passing with an equivariance order of $l=2$, and is trained on DFT relaxation trajectories\cite{deng2023chgnet} from the Materials Project dataset. Similarly, the medium  MACE-MPA-0 has a message-passing equivariance order of $l=1$, and is trained on the Alexandria\cite{schmidt2024improving} and Materials Project dataset.
    \item MatterSim v1 5M (MatterSim)\cite{yang2024mattersim}: The MatterSim employs three-body interactions and atomic positions, similar to the pioneering MLIP M3GNet architecture. MatterSim is pre-trained on the proprietary MatterSim dataset.
    \item DPA3-v1-OpenLAM (DPA3): The DPA3 model is a message passing large atomic model\cite{Zhang2024}, which is pretrained on OpenLAM dataset \cite{peng2025openlamchallenges}, a diverse interdisciplinary collection of datasets, and fine-tuned on the Alexandria and Materials Project datasets.
    \item ORB v2 (ORB)\cite{neumann2024orbfastscalableneural}: The ORB model is notable for combining a graph network simulator with smooth overlap of atomic positions. Additionally, forces in the ORB model are predicted separately from the energy surface, hence, they do not represent the energy gradient.  ORB is pre-trained on DFT relaxation trajectories from the Materials Project and the Alexandria datasets.
    \item GRACE-2L-OAM (GRACE)\cite{bochkarev2024graph}: The GRACE model utilizes a graph atomic cluster expansion, enabling efficient descriptions of semilocal interactions\cite{bochkarev2024graph}. The two-layer semi-local GRACE-2L-OAM model is pre-fitted on the OMat24 dataset\cite{barrosoluque2024omat24}, and fine-tuned on the Alexandria and Materials Project datasets.
\end{itemize}

\section{Results}
In this section, we present the results of our benchmarking of the six dispersion-corrected MLIPs on the vdW heterobilayer dataset. We first consider performance in terms of relaxed atomic structures and subsequently investigate, for two selected MLIPs, how these differences propagate to the electronic band structures. Finally, we describe the 2D-vdW heterostructure database HetDB. 

\subsection{Atomic structures}
\begin{figure}
    \centering
    \includegraphics[width=\linewidth]{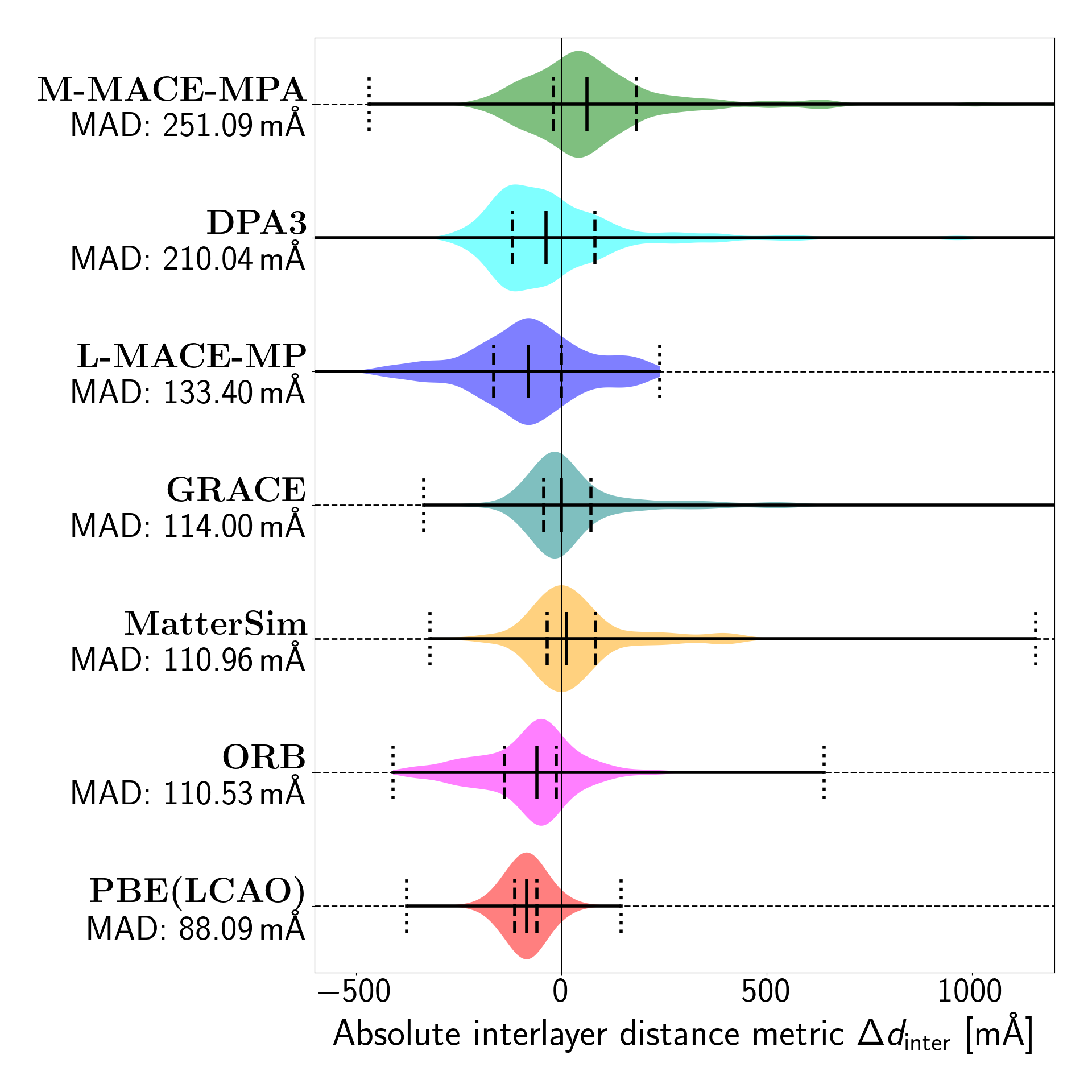}
    \caption{Violin plot showing the distribution of the error on the interlayer distance, Eq. (\ref{eq:interlayer_delta}), for the 336 vdW heterostructures relaxed with six different dispersion-corrected MLIPs and PBE-D3(LCAO). The deviations are measured relative to the PBE-D3(PW) interlayer distance, which represents the ground truth for all the models. The full, dashed, and dotted lines represent every 25th percentile starting from zero, where the full line is the median.}
    \label{fig:z-violin}
\end{figure}

\begin{figure}
    \centering
    \includegraphics[width=\linewidth]{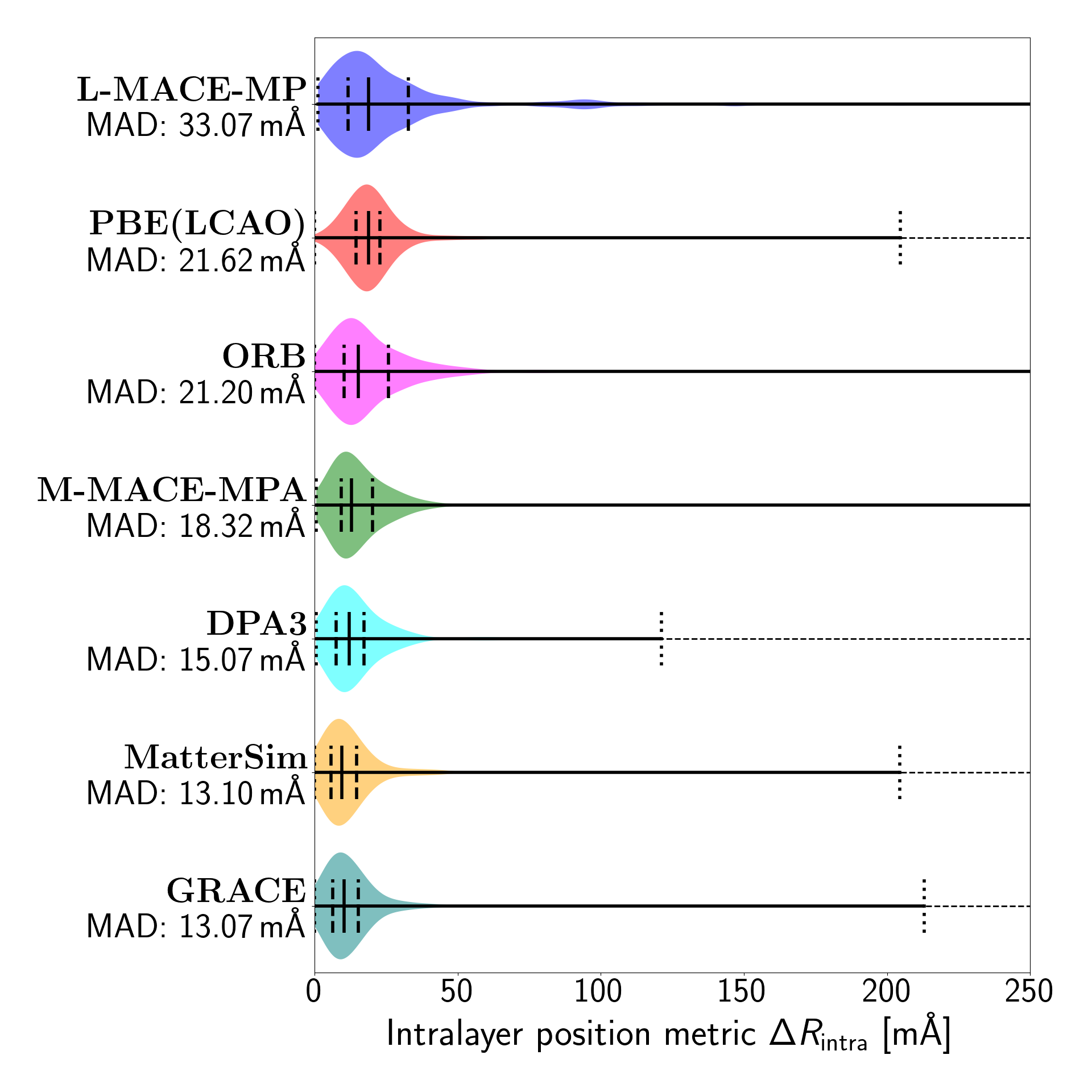}
    \caption{Violin plot showing the distribution of the intralayer metric, Eq. (\ref{eq:intra}), for the 336 heterostructures. The metric quantifies the change in the atomic structure of the individual monolayers of the heterostructures relative to the PBE-D3(PW) ground truth. The full, dashed, and dotted lines represent every 25th percentile starting from zero, where the full line is the median.}
    \label{fig:rmsd-violin}
\end{figure}

\begin{figure}
    \centering
    \includegraphics[width=\linewidth]{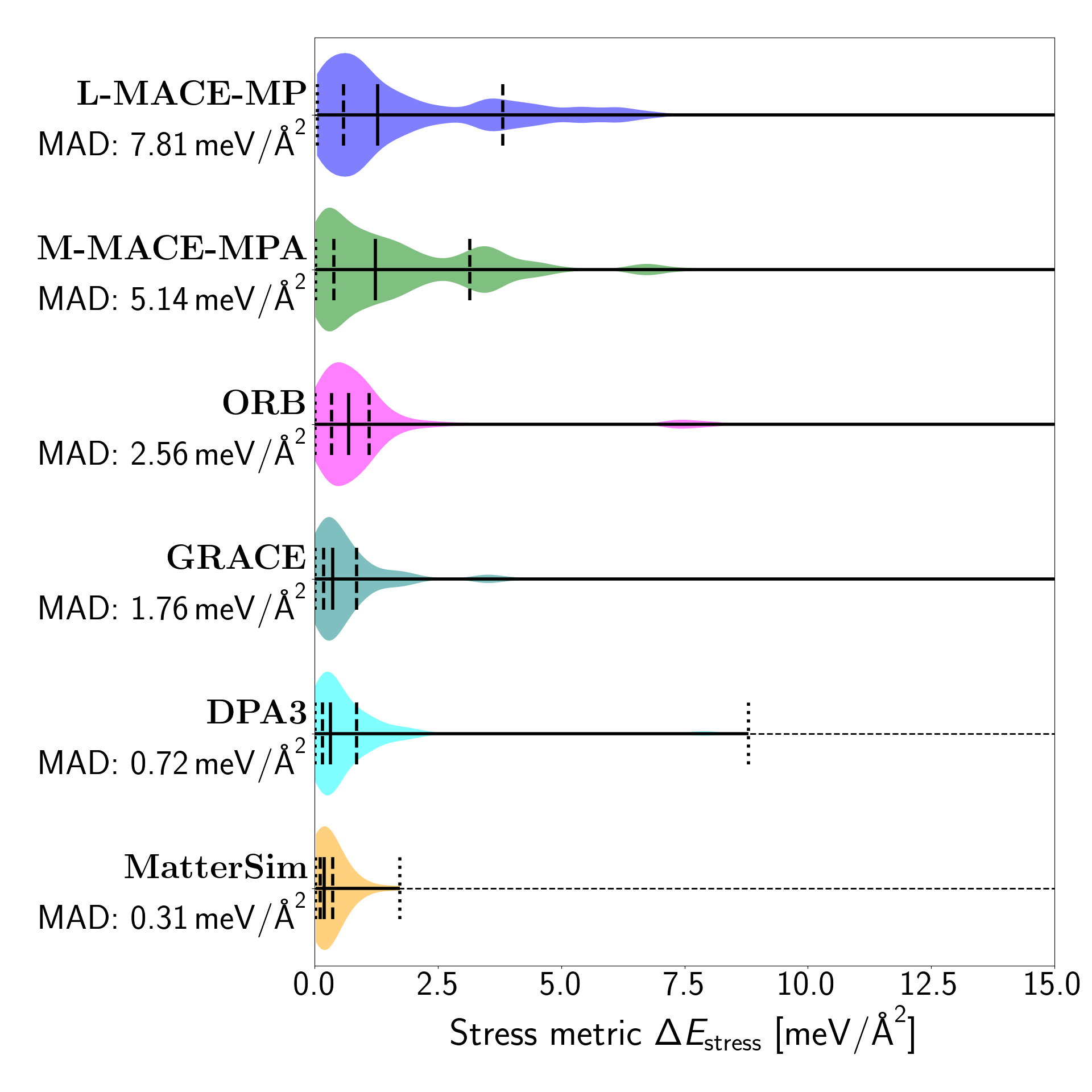}
    \caption{Violin plot showing the distribution of the stress metric, Eq. (\ref{eq:stress_metric}), quantifying the mechanical energy due to internal strain in the heterostructures. The full, dashed, and dotted lines represent every 25th percentile starting from zero, where the full line is the median.}
    \label{fig:epot-violin}
\end{figure}

\begin{figure}[ht]
    \centering
    \includegraphics[width=\linewidth]{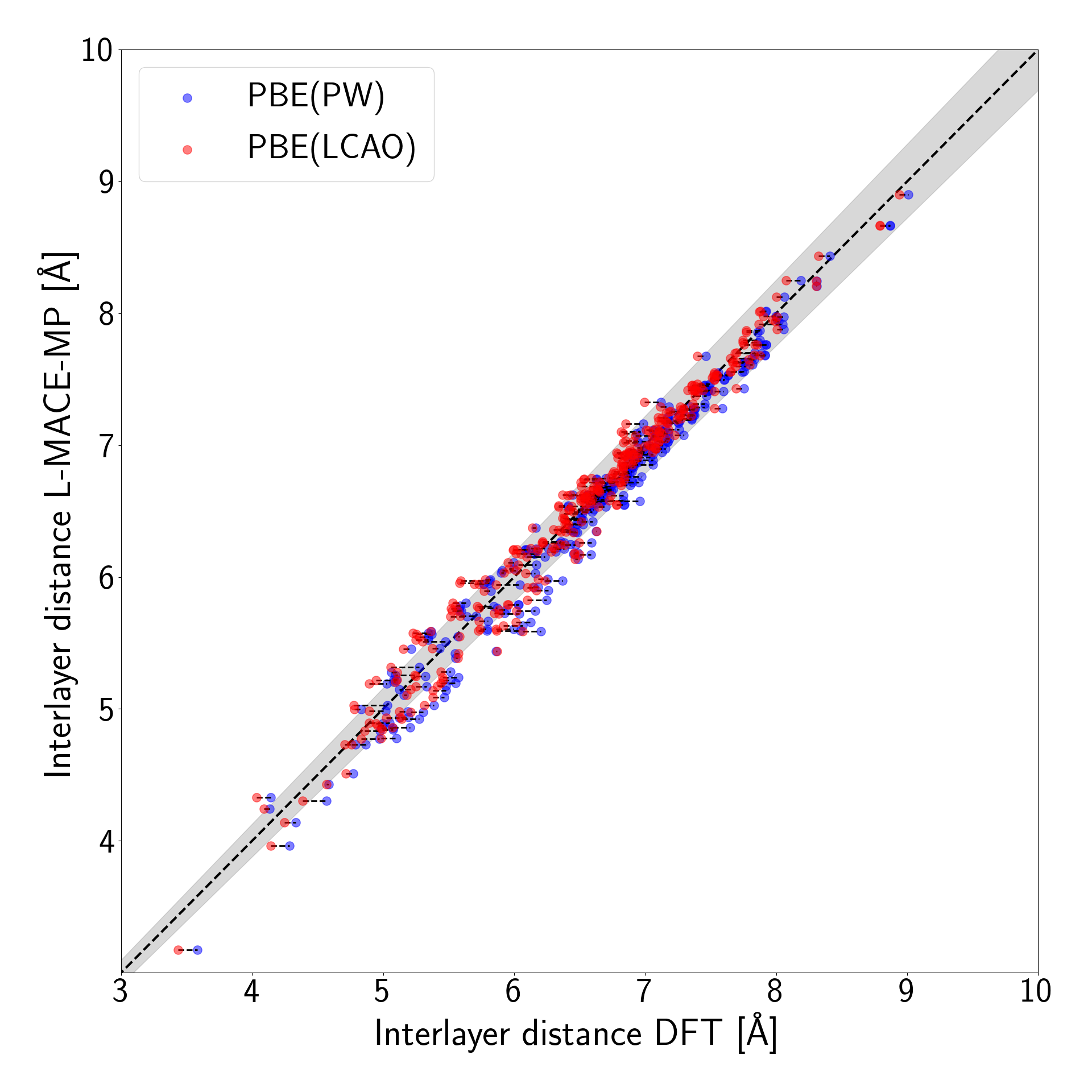}
    \caption{Average interlayer distance, Eq. (\ref{eq:avg_distance}), for the 336 heterostructures relaxed using the L-MACE-MP versus 
    PBE(PW) and PBE(LCAO), respectively. In all cases, dispersive interactions are included by the D3 correction. The grey shaded region represents the expected uncertainty ($\pm 3.1\%$) on the DFT-calculated interlayer distances (with a converged PW basis) stemming from the xc-functional\cite{tran2019nonlocal} (see text).}
    \label{fig:z-stats}
\end{figure}

Figure \ref{fig:z-violin} shows the error distribution of the interlayer distances obtained with the different MLIPs. The best performance is seen for ORB, MatterSim, and GRACE, which all yield mean absolute deviations (MAD) close to $0.11 \, \text{Å}$, not far from the DFT result with the LCAO-DZP basis set ($0.09 \, \text{Å}$). The DPA3 and the M-MACE-MPA models both underperform with MADs of $0.21 \, \text{Å}$ and $0.25 \, \text{Å}$, respectively. Table 2 in the supplementary information (SI) shows a direct comparison between all the models. From the distributions in Fig. \ref{fig:z-violin} and percentiles (indicated by vertical lines) it follows that MatterSim and GRACE have rather symmetric error distributions around zero and thus no systematic errors. In contrast, ORB, DPA3, L-MACE-MP, and PBE(LCAO) all systematically underestimate the interlayer distance while M-MACE-MPA overestimates it. Interestingly, in terms of outliers, PBE(LCAO), L-MACE-MP, and ORB perform the most consistent, with maximum outliers of $0.38 \, \text{Å}$, $0.61 \, \text{Å}$, and $0.64 \, \text{Å}$, respectively, as opposed to the least consistent models M-MACE-MPA and DPA3 with maximum outliers of $2.07 \, \text{Å}$ and $2.34 \, \text{Å}$, accordingly.

Next, we consider the ability of the MLIPs to describe the internal atomic structure of the monolayers in the heterostructures. The intralayer metric in Eq. (\ref{eq:intra}) is designed to measure just that. Figure \ref{fig:rmsd-violin} shows the distribution of the RMSD of the intralayer atomic positions relative to the PBE-D3(PW) ground truth after relaxation with each MLIP. MatterSim and GRACE perform the best with mean absolute deviation (MAD) over all the heterostructures of $13 \, \text{mÅ}$. They are followed by DPA3 with a MAD of $15 \, \text{mÅ}$. With a MAD of $18 \, \text{mÅ}$ the medium MACE model is significantly better than the large MACE model, which produces a MAD of $33 \, \text{mÅ}$ making it the least accurate model. The good performance of the MLIPs on the intralayer metric is evidenced by the fact that, except for L-MACE-MP, all the models perform better than the PBE-D3(LCAO) method, which achieves a MAD of $21 \, \text{mÅ}$. Additionally, we observe that while MatterSim and Grace perform the best in terms of MAD, DPA3 behaves the most consistent, with the lowest maximum outlier of $121 \, \text{mÅ}$.

All the calculations employ a moiré supercell defined from the PBE-D3(PW) monolayer lattice constants, see section \ref{sec:moire}. This supercell is kept fixed during the relaxation. While the moiré cell by construction yields low strain/stress in the PBE-D3(PW) calculation, this may not be the case for all the MLIPs. To measure the amount of internal stress relative to the PBE-D3(PW) ground truth value, we use the stress metric defined in Eq. (\ref{eq:stress_metric}). Figure \ref{fig:epot-violin} shows the distribution of the stress metric across the 336 heterostructures. With a MAD of $0.31 \, \text{meV}/\text{Å}^2$ and maximum outlier of $1.73 \, \text{meV}/\text{Å}^2$, MatterSim is by far the best performing model followed by DPA3 ($0.72 \, \text{meV}/\text{Å}^2$ MAD), GRACE ($1.76  \, \text{meV}/\text{Å}^2$ MAD), and ORB ($2.58  \, \text{meV}/\text{Å}^2$ MAD), respectively. In contrast, the two MACE models shows poor performance with MADs of above $5 \, \text{meV}/\text{Å}^2$. 
We note that the PBE-D3(LCAO) result is not included in Figure \ref{fig:epot-violin} because the stress tensor is not available in GPAW in LCAO mode. 

In the rest of the paper we specialize to the L-MACE-MP and MatterSim models. 

Figure \ref{fig:z-stats} shows the interlayer distance, Eq. (\ref{eq:avg_distance}), in the 336 heterobilayers after relaxation with L-MACE-MP ($y$-axis) and DFT-PBE ($x$-axis) with either a PW basis set (blue circles) or LCAO basis set (red circles). In all calculations dispersion forces are included by the D3 correction.

The grey shaded area on the plots ($\pm 3.1\%$) represents the expected uncertainty on the DFT-calculated interlayer distances (for a fully converged PW basis) stemming from the exchange-correlation (xc-)functional. This value has been obtained as the standard deviation of the interlayer distances calculated with 13 different non-local xc-functionals for a set of ten vdW bulk crystals\cite{tran2019nonlocal}. In comparison, the mean absolute relative deviation (MARD) between the PBE-D3 and MACE-D3 interlayer distances is 2.07\% and 2.40\% for the PW and LCAO basis set, respectively (see Table 2 in the supplementary information). Both values are comparable to (in fact smaller than) the uncertainty stemming from the xc-functional. As the latter can be regarded as an intrinsic DFT uncertainty we can conclude that the MLIP reaches DFT accuracy for the interlayer distance.

We note in passing that the interlayer distances predicted by the LCAO basis set are systematically smaller (by $\sim 0.1$ \AA) than those predicted by the PW basis set. A likely explanation is that the LCAO basis functions have finite range, and therefore, the effect of Pauli repulsion\cite{rusu2010first,gong2010first} due to overlap of the wave functions from the two layers, is underestimated with the LCAO basis. Table 3 in the supplementary information summarizes the mean deviations between the interlayer distances calculated with the three different methods. 


\subsection{Electronic structure}

\begin{figure}[ht]
    \centering
    \includegraphics[width=.95\linewidth]{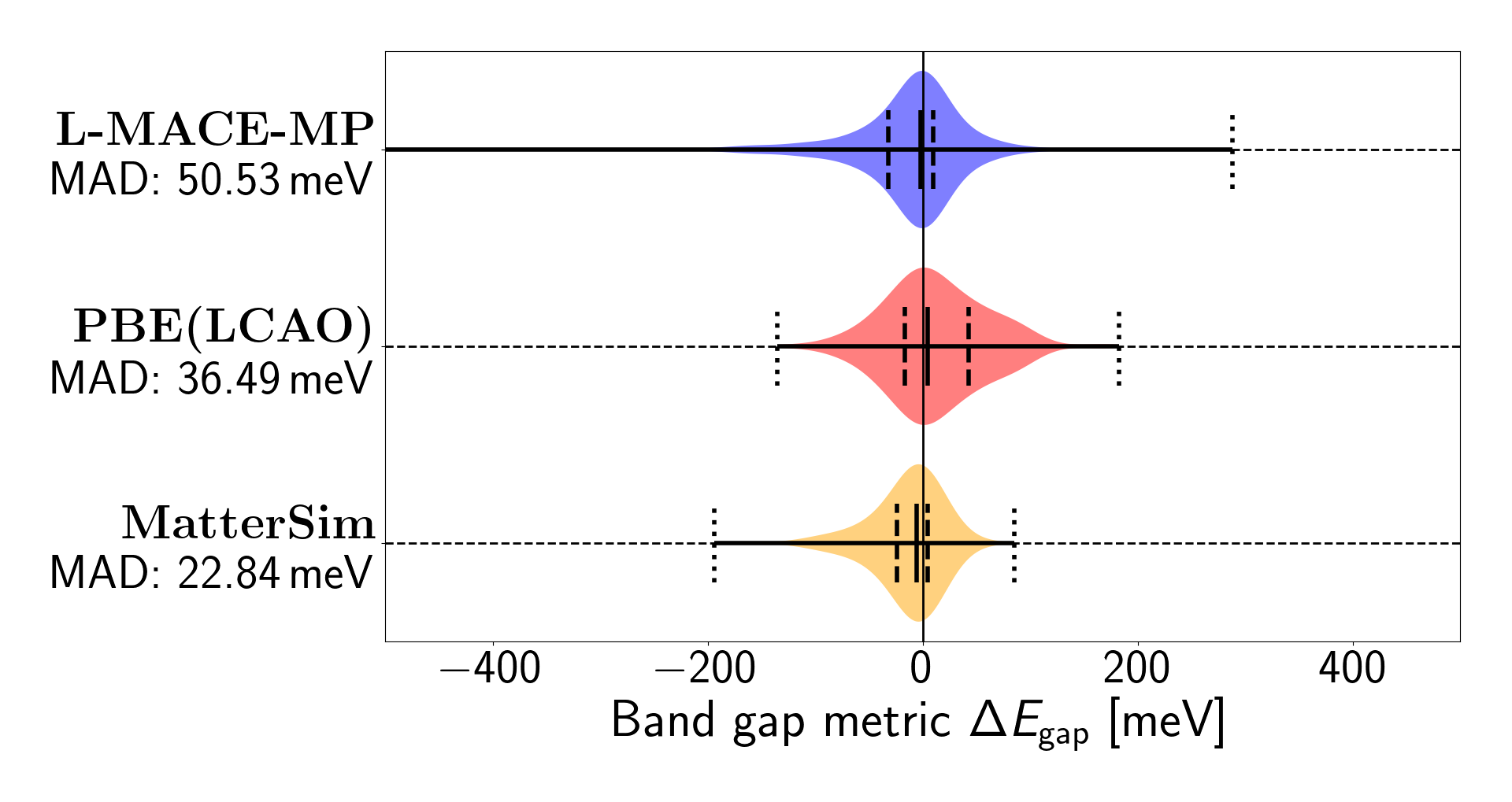}
    \caption{Violin plot of the electronic band gap differences for heterostructures relaxed using MatterSim, L-MACE-MP0, and PBE(LCAO) relative to the PBE(PW) ground truth, see Eq. (\ref{eq:gap_metric}). For all structures, the band gap has been calculated by PBE(LCAO). The full, dashed, and dotted lines represent every 25th percentile starting from zero, where the full line is the median.}
    \label{fig:gap_violin}
\end{figure}

\begin{figure}[ht]
    \centering
    \includegraphics[width=.95\linewidth]{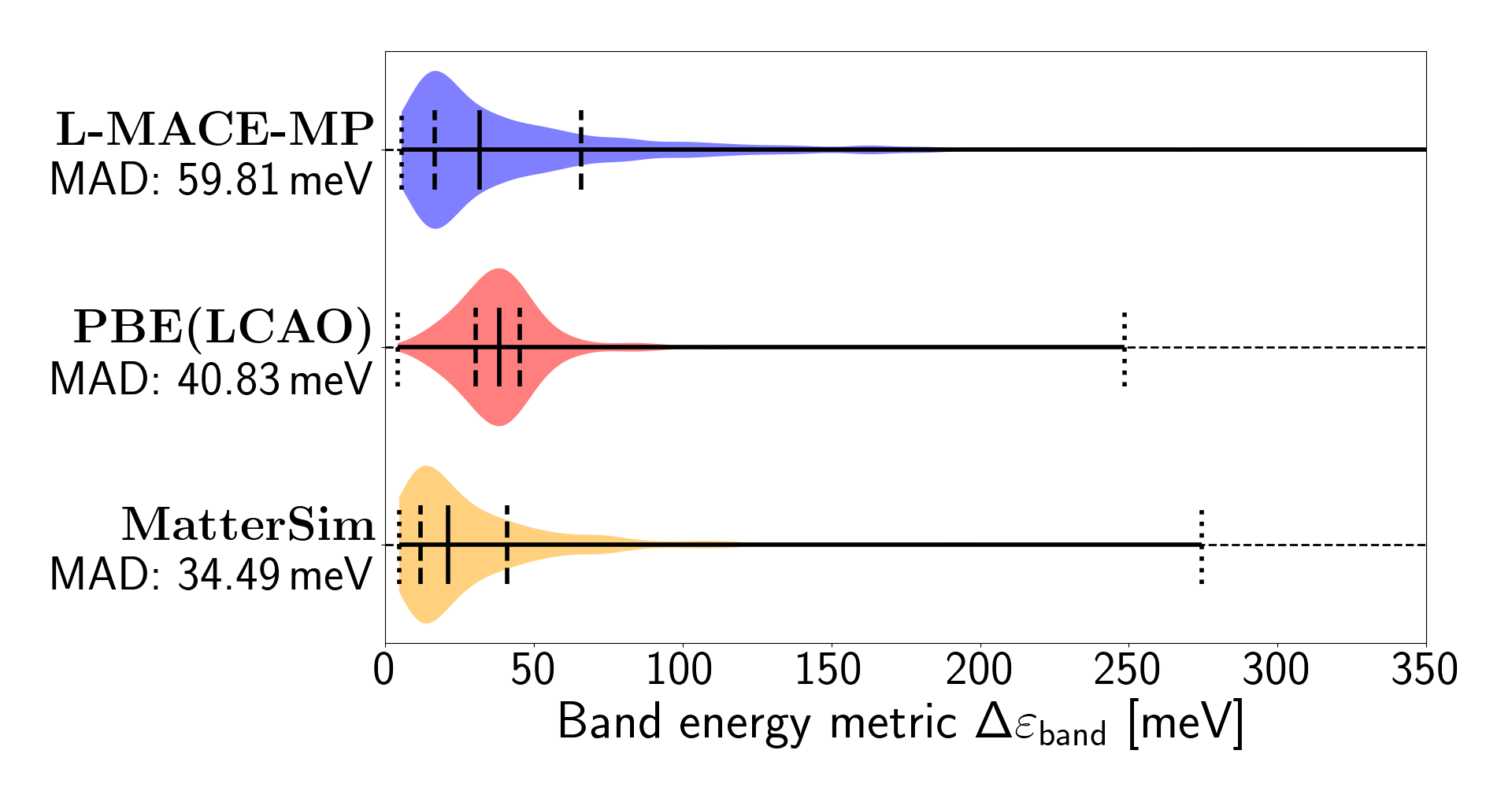}
    \caption{Violin plot of the electronic band structure differences for heterostructures relaxed using MatterSim, L-MACE-MP0, and PBE(LCAO) relative to the PBE(PW) ground truth, see Eq. (\ref{eq:band_metric}). For all structures, the band structure has been calculated by PBE(LCAO). The full, dashed, and dotted lines represent every 25th percentile starting from zero, where the full line is the median.}
    \label{fig:comps-rmsd}
\end{figure}

In the previous section, we found that MACE-D3 yields interlayer distances of vdW heterobilayers within the DFT uncertainty of nonlocal xc-functionals. Next, we investigate the single-particle band structures of the bilayers obtained after relaxations with MACE-D3 and PBE-D3, respectively. This is relevant because most of the interest in vdW heterostructures originates from their electronic, optical, or magnetic properties. To judge the relevance of a MLIP for a given application, it is therefore crucial to know how the structural inaccuracies affect the electronic structure.  

Figure \ref{fig:gap_violin} shows the distribution of the single-particle band gap of the 336 heterobilayers. All band structures are calculated with the PBE xc-functional and the LCAO-DZP basis set, but using atomic structures obtained from relaxations with (dispersion-corrected) L-MACE-MP, MatterSim, PBE(LCAO), and PBE(PW), respectively. Thus, the only difference between the results stems from differences in the atomic structures. 

The band gaps calculated for the MatterSim relaxed structures have a MAD of $22.8 \, \text{meV}$. This is an excellent accuracy and even better than obtained for the PBE(LCAO) relaxed structures ($36.5 \, \text{meV}$), while showing similar consistency in terms of outliers. With a MAD of $50.5 \, \text{meV}$, the performance of the L-MACE-MP appears less impressive. However, from the distributions in Figure \ref{fig:gap_violin} it is clear that the reason for the larger MAD obtained for MACE is the presence of a few significant outliers, while the error distribution for the vast majority of systems is very similar to that of MatterSim.

Figure \ref{fig:comps-rmsd} shows the distribution of the band energy metric, defined as the RMSD of the full band structure energies relative to the PBE-D3(PW) ground truth. Again, the overall performance of the MLIPs is excellent. The trends are the same as seen for the band gap energy: MatterSim reaches the highest accuracy (MAD of $34.5 \, \text{meV}$) while L-MACE-MP produces a very similar error distribution but with a few significant outliers resulting in a larger MAD of $59.8 \, \text{meV}$. For comparison, the PBE-D3(LCAO) result achieves an intermediate MAD of $40.8 \, \text{meV}$, with maximal outliers on the same order of MatterSim.

\subsection{The vdW heterostructure database: HetDB}
\begin{figure*}[ht]
    \centering
    \includegraphics[width=\linewidth]{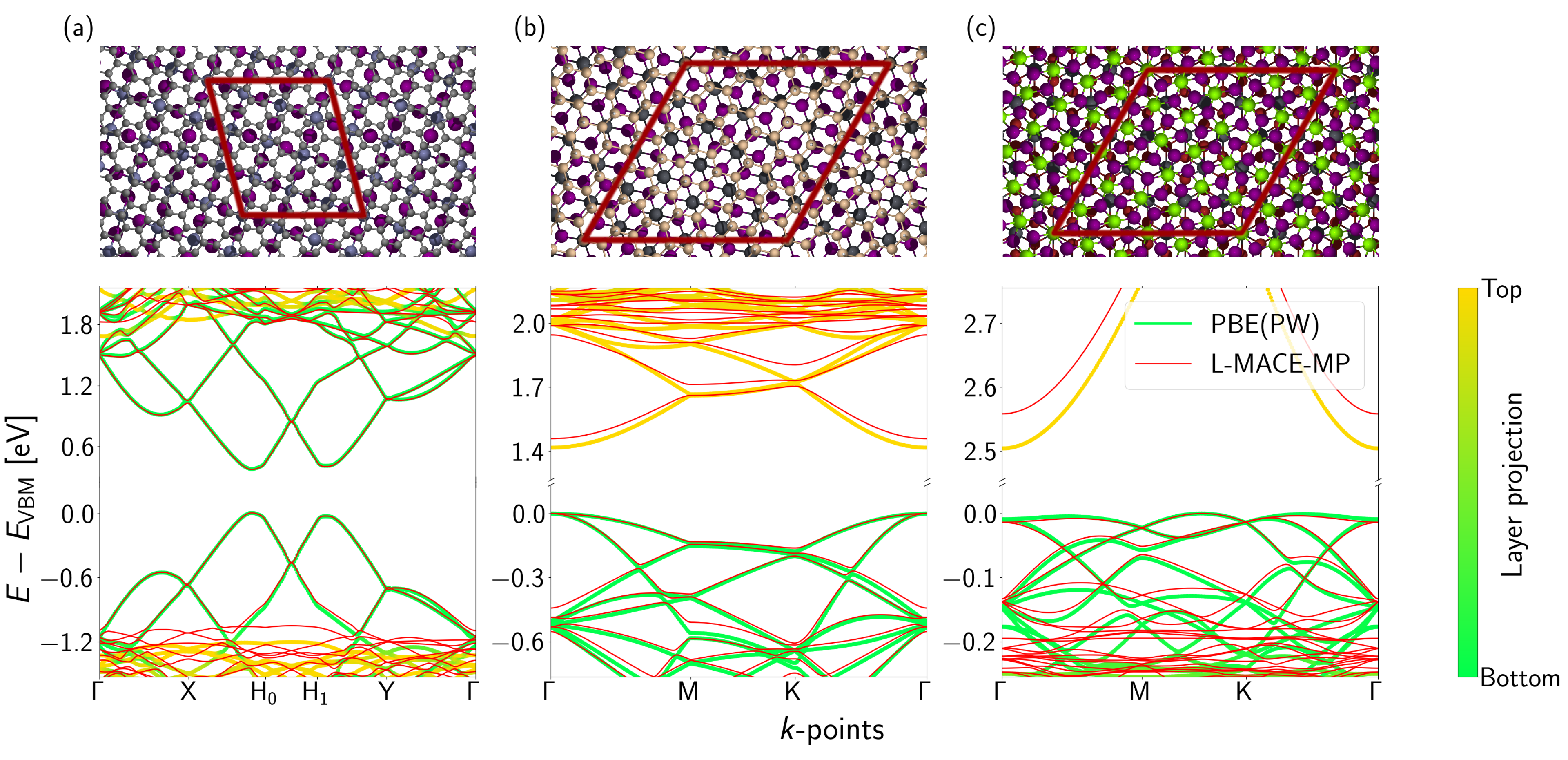}
    \caption{Examples of heterobilayers and their electronic band structures. The three structures are (a) Graphene-ZnI$_2$, (b) Si$_2$H$_2$-PbI$_2$, and (c) MgI$_2$-PbBr$_2$. The moiré cell is indicated in red on the atomic structures. For all structures, the band structure has been calculated by PBE(LCAO). The electronic band structures corresponding to the heterostructures relaxed with PBE-D3(PW) (colored lines) are projected onto the top and bottom layers. The band structure calculated for the structure relaxed with L-MACE-MP (red line) is also shown. }
    \label{fig:bsfig}
\end{figure*}

The atomic structures of the 336 heterobilayers relaxed with PBE-D3(PW) as well as the calculated electronic band structures are available for download or online browsing via the HetDB database\cite{hetdb}. The band structures are calculated with PBE(LCAO) and include spin-orbit interactions and projections onto the top and bottom layers (see below). Importantly, the HetDB is seamlessly integrated with the C2DB monolayer database\cite{article:C2DB} and the BiDB homobilayer database\cite{pakdel2024high} allowing for direct and easy comparisons of mono- and bilayer properties. We are planning to expand the HetDB continuously, adding more heterostructures and physical properties.  

To illustrate the type of information currently available in HetDB, Fig. \ref{fig:bsfig} shows three examples of electronic band structures (for simplicity, spin-orbit coupling is not included in these examples). Because the main focus of the current work is on MLIP benchmarking, we compare the band structures of bilayers optimized with PBE-D3(PW) and the large MACE-MP model with the D3 correction, respectively. For the reference bilayer structure relaxed with PBE-D3(PW), the bands are projected onto the top and bottom layers as encoded by the color of the bands. Qualitatively, we find good agreement between the band energies of the two bilayer structures. However, there are some noticeable discrepancies, in particular away from the valence bands, as also expected from the MAD of the band energy metric for the L-MACE-MP model.

Figure \ref{fig:bsfig}(a) shows the band structure of a Graphene-ZnI$_2$ bilayer containing 105 atoms in the moiré cell. The bilayer exhibits a direct intralayer gap along the symmetry path of $\sim 0.3 \, \text{eV}$ localized in the graphene layer. We note that this band gap is larger than the true band gap obtained when considering the band structure over the full 2D Brillouin zone, which yields a tiny direct gap of $\sim 20 \, \text{meV}$. This pronounced difference occurs because the band path (defined from the moiré cell of the bilayer) does not pass through the region corresponding to the $K$-point in the Brillouin zone of the isolated graphene sheet, which is where the band gap is located. 

Fig. \ref{fig:bsfig}(b) shows the band structure of a Si$_2$H$_2$-PbI$_2$ bilayer with 240 atoms in the moiré cell. This structure features a direct interlayer band gap of 1.5 eV at the $\Gamma$-point, making it an interesting candidate for realizing long-lived, tunable interlayer excitons in the optical regime\cite{peimyoo2021electrical}


Figure \ref{fig:bsfig}(c) shows the band structure of a MgI$_2$-PbBr$_2$ bilayer containing 159 atoms in the moiré cell. This band structure features an indirect interlayer band gap of around 2.5 eV. Interestingly, the valence bands localized on the bottom layer are very flat compared to the conduction band localized on the top layer, which are much more dispersive. This different nature of the conduction and valence bands is presumably a result of the interlayer character of the band structure, which implies that the two bands live on distinct materials.   

\section{Discussion}
We have benchmarked six machine learning interatomic potentials (MLIPs) augmented with the D3 dispersion correction on a set of 336 non-magnetic and non-metallic van der Waals heterostructures containing between 4 and 300 atoms in the unit cell. Through comprehensive benchmarking of both structural and electronic properties we have demonstrated that some of the MLIPs reach accuracies comparable to the uncertainty on density functional theory calculations stemming from the choice exchange-correlation functional. Out of the six benchmarked machine learning models, MatterSim achieves the best performance with errors that are mostly lower than those achieved by DFT when employing a common atomic orbital basis set. The average errors on the electronic band structures that derive from the structural inaccuracies produced by MatterSim are as low as 23 meV for the band gap and 35 meV for general band energies. These errors are significantly lower than both the numerical precision of state of the art GW band structure calculations of solids\cite{azizi2025precision} and the accuracy of such calculations relative to experimental values\cite{van2006quasiparticle}. 
On this basis, we conclude that MLIPs with dispersion corrections are ready for large-scale structural simulations of 2D vdW materials, at least for the important class of non-magnetic and non-metallic materials considered in this work. All the results from the current study are available in a curated database, HetDB, which we plan to develop further both in terms of structures and properties. We believe that the HetDB will be useful as a platform for exploring and designing vdW heterostructures with specific physical properties and as a data resource for future development and benchmarking of machine learning models targeting complex vdW materials.       

\section{Methods}
\label{sec:methods}
Figure \ref{fig:intro} summarizes the workflow used to select the monolayers to be stacked, define the moiré supercells, relax the heterostructures, and calculate their electronic band structures. The computational workflow was managed using the MyQueue task scheduler\cite{mortensen2020myqueue}. Below, we describe each step in more detail.

\subsection{Selection of monolayers}
The monolayers used to construct the heterobilayers are obtained from the C2DB\cite{article:C2DB}. We search for monolayers with less than seven atoms in the primitive unit cell that are non-magnetic, dynamically stable, non-metallic, and have high thermodynamic stability expressed by an energy above the convex hull, $E_{\text{hull}}$ below 10 meV/atom. Furthermore, we require that the GW band structure is available in C2DB and
that either the valence band maximum (VBM) or conduction band minimum
(CBM) is located at the $\Gamma$-point. These requirements
are in fact of little relevance to the current work, but will be central to a future study. Additionally, we include the monolayers: Graphene, hexagonal boron-nitride (h-BN), MoS$_2$, MoSe$_2$, MoTe$_2$, WS$_2$,
WSe$_2$, and WTe$_2$ due to their high relevance in the field of 2D materials. The resulting 44 monolayers are relaxed using the Perdew-Burke-Ernzerhof (PBE) \cite{perdew1996generalized} xc-functional with the D3 dispersion correction\cite{grimme2010consistent}, and the moiré cells are constructed as described below.

\subsection{Finding the moiré supercells}\label{sec:moire}
For each combination of the 44 selected monolayers we check whether a moiré supercell, represented as integer linear combinations of the PBE-D3 monolayer cell vectors, can be constructed such as to satisfy the following two conditions: (i) the strain on both monolayers is below 1\% and (ii) the number of atoms in the supercell is below 300. If several such cells are found, we pick the one with the fewest number of atoms. More details on the procedure used to determine the supercell can be found in Appendix \ref{app:findmoire}.

The above procedure yields supercells in which the two monolayers are under a small but finite strain. The strain of the two monolayers, $\sigma_1$ and $\sigma_2$, are functions of the supercell basis vectors. We fine-tune the latter by minimizing the mechanical energy 
\begin{equation}\label{eq:mech}
E_{\mathrm{stress}} = \frac{1}{2}(\sigma_1 S_1 \sigma_1+\sigma_2 S_2 \sigma_2),
\end{equation}
where $S_i$ is the stiffness tensor of layer $i$. 

In total, we generate 336 heterobilayers satisfying the stated conditions. Table 1 in the supplementary information provides an overview of all the heterobilayers including the layer groups (the 2D analogue of the space group\cite{fu2024symmetry}), band gaps of the bilayer and the constituent monolayers, the in-plane strain of each monolayer, the twist angle, and the number of atoms in the moiré cell. The data can also be browsed online at \cite{hetdb}.

\subsection{Relaxation and band structure calculations}
The monolayers used to build the heterostructures were relaxed with PBE-D3 using a plane wave (PW) basis with a cutoff energy of $800 \, \text{eV}$. Structural relaxation was continued until the maximum force was below $10 \, \text{meV/Å}$. The calculations utilized a $\Gamma$-centered $k$-point grid with a density of $6 \, \text{Å}$ and a Fermi-Dirac occupation smearing with a width of $0.05 \, \text{eV}$. Additionally, the stiffness tensor was evaluated for the monolayers using strains of $0.1 \%$ (in both $xx$, $yy$, and $xy$ directions). The stiffness tensor is used by our algorithm to find the zero-stress supercells of the heterostructure.

Before relaxing the atomic positions of the heterostructure, an initial guess for the interlayer distance is found by minimizing the total energy as function of the distance between layers while keeping the atoms fixed in the relaxed monolayer configuration. In detail, the "$z$-scan algorithm" evaluates the total energy at 25 evenly spaced points in the layer distance interval $2-5\text{Å}$, with additional points being allocated outside this region when needed. The binding energy curve is interpolated using cubic splines before the minimum is determined. The total energies are calculated using various methods: The PBE-D3 xc-functional with a basis consisting of plane waves (PW) or linear combination of numerical atomic orbitals (LCAO) basis, as well as the MLIP-D3 models. Once an optimal initial interlayer distance has been found, the atomic positions are relaxed to a maximum force of $50 \, \text{meV/Å}$. For both the $z$-scan and atomic relaxations, the DFT calculations were performed with a $\Gamma$-centered $k$-point grid with a density of $6 \, \text{Å}$, and a Fermi-Dirac occupation smearing with a width of $0.05 \, \text{eV}$. For the PW calculations, a cutoff energy of $600 \, \text{eV}$ was applied, while the double-zeta with polarisation (DZP) basis set was utilized for LCAO\cite{larsen2009localized}. All DFT calculations were performed with the GPAW code\cite{mortensen2024gpaw}. 



All calculations apply zero-damping for the D3 dispersion correction, as well a cutoff radius of $50.2 \, \text{Å}$ for pair interactions and $20 \, \text{Å}$ for coordination numbers.

Finally, the electronic band structure is calculated using PBE with the LCAO-DZP basis set for all the heterobilayers. To assess the influence of the structural variations on the band energies, the latter are calculated for the heterobilayers obtained after relaxation with DFT-PBE-D3(PW), DFT-PBE-D3(LCAO), and the D3-corrected models MatterSim and L-MACE-MP. In all cases, the band structures are calculated on top of a well converged ground state density obtained with a $k$-point density of $24 \, \text{Å}$ and a Fermi-Dirac occupation smearing with a width of $0.05 \, \text{eV}$.

\section{Acknowledgement}
The authors acknowledge funding from the Villum Investigator Grant No. 37789 supported by VILLUM FONDEN and from the Novo Nordisk Foundation Data Science Research Infrastructure 2022 Grant:  A high-performance computing infrastructure for data-driven research on sustainable energy materials, Grant no. NNF22OC0078009.

\appendix
\onecolumngrid
\section{Finding moiré supercells}
\label{app:findmoire}
The moiré supercells of the 336 heterobilayers are found following the algorithm described below. First we define the unstrained supercell lattice vectors for the two layers as
\begin{align}
    R_{l,v}^{(a)} &= c_{l,c}^{(a)} r_{c,v}^{(a)}, \\
    R_{l,v}^{(b)} &= c_{l,c}^{(b)} r_{c,v}^{(b)},
\end{align}
where $l$ is the lattice vector index, $a, b$ bottom and top layer, $c_{l,c}^{(n)}$ coefficients for linear combinations of $r_{c,v}^{(n)}$ the lattice vectors of the primitive cell for layer $n$. From these unmatched unstrained supercell lattices of each layer, the algorithm finds the stress minimized common supercell, using the elasticity tensor, by varying the common supercell lattice $\bar{R}_{l,v}$. Finally, the strain is calculated for each layer as the finite Lagrangian strain tensor $S_{v,v}^{(n)}$, given by the expression \cite{Schlenker:a15471}
\begin{align}
    s_{v,v}^{(n)} &= ({R}_{l,v}^{(n)})^{-1} \bar{R}_{l,v} - I_{v,v}, \\
    \label{eq:straintensor}
    S_{v,v}^{(n)} &= \frac{1}{2} \left( s_{v,v}^{(n)} + (s_{v,v}^{(n)})^T + s_{v,v}^{(n)} (s_{v,v}^{(n)})^T \right),
\end{align}
where $I_{v, v}$ is the identity matrix. From this, the maximum strain is given as $\text{max}_{n,v} (\lambda_v^{(n)})$ with $\lambda_v^{(n)}$ being the eigenvalues of $S_{v,v}^{(n)}$.
Here we summarize the main conditions used to construct the bilayers: 
\begin{itemize}
    \item The maximum allowed strain of the individual monolayers is $S_{\text{max}} = 1\%$.
    \item The twist angle (i.e. angle that the top layer is rotated compared to the bottom layer) is within $\phi_b = [0^\circ, 90^\circ]$.
    \item The number of atoms in the supercell is as small as possible, with the maximum number allowed being $N_{\text{max}} = 300$.
    \item The internal angle (i.e. angle between the supercell lattice vectors) is in the interval $\chi_b = [15^\circ, 165^\circ]$.
    \item The maximum allowed ratio between the lattice vector norms of the supercell is $R_{\text{max}} = 10$.
    \item All coefficients for the linear combinations of primitive lattice vectors are less than $c_{\text{max}} = 25$.
\end{itemize}
The first three conditions (strain, twist angle, and number of atoms) are physical constraints on the lattices we search for, while the last three (internal angle, norm ratio, and maximum coefficients) are numerical constraints, necessary for reducing the otherwise infinite linear combinations.
A more detailed description of the algorithm follows below (written in Einstein sum notation):
\begin{itemize}
    \item Generate all possible linear combinations of the lattice vectors of each layer $n$ (henceforth supercell lattice vectors $R^{(n)}_{lv} = c_{lc} r^{(n)}_{cv}$, with coefficients $\big\{c_{lc} \in \mathbb{Z} \, \big| \, \abs{c_{lc}} < c_{\text{max}} \land (c_{l0}, c_{l1}) \neq (0, 0) \big\}$.
    \item Generate matching pairs of supercell lattice vectors from the two layers $S_{\lambda} = (R^{(a)}_{lv}, R^{(b)}_{mv})$ with $\lambda = (l, m)$, where:
    \begin{itemize}
        \item Approximate uniaxial strain is low $\norm{R^{(a)}_{lv} - R^{(b)}_{mv}}_v / \norm{\bar{R}_{\lambda v}}_v < 1.5 S_{\text{max}}$, with $a, b$ being the two separate layers, and $\bar{R}_{\lambda v}$ being the approximate unstressed cell vector assuming only uniaxial strain along one lattice vector direction.
        \item The approximate twist angle $\phi_{\lambda} = (\text{Angle}(R^{(b)}_{mv}) - \text{Angle}(R^{(a)}_{lv}) \mod 2 \pi) \in \phi_b$, where $\text{Angle}(x)$ operates on the last dimension of its input and returns the angle of the vector.
    \end{itemize}
    \item Construct supercells for the two layers from pairs of the selected supercell lattice vectors $(S_{\lambda}, S_{\mu})$, s.t. $\rho^{(a)}_{X \alpha v} = \left[ R^{(a)}_{\lambda v}, R^{(a)}_{\mu v} \right]$ and $\rho^{(b)}_{X \alpha v} = \left[R^{(b)}_{\lambda v}, R^{(b)}_{\mu v} \right]$ define lists of supercells, with $X = (\lambda, \mu)$ indexing the moiré superlattice. These moiré superlattices are filtered according to
    \begin{itemize}
        \item The difference in approximate twist angle $\abs{\phi_{\lambda} - \phi_{\mu}} < 1.2 \pi S_{\text{max}}$ radians.
        \item The number of atoms in the supercell $\sum_n N_n \text{Area}(\rho^{(n)}_{X \alpha v}) / \text{Area}(r^{(n)}_{cv}) \leq N_{\text{max}}$, where $\text{Area}(x)$ operates on the last two dimension of its input and returns the cell area.
        \item The internal angle of both layers supercell $\abs{\text{Angle}(\rho^{(n)}_{X 0 v}) - \text{Angle}(\rho^{(n)}_{X 1 v})} \in \chi_b$.
        \item The norm ratio $\text{Max}(\norm{\rho^{(n)}_{X 0 v}}_v / \norm{\rho^{(n)}_{X 1 v}}_v, \norm{\rho^{(n)}_{X 1 v}}_v / \norm{\rho^{(n)}_{X 0 v}}_v) \leq R_{\text{max}}$.
    \end{itemize}
    \item Remove non-unique supercells. This is done by calculating the deformation tensor between the supercells of the two layers, $s_{Xvv} = (\rho_{X \alpha v}^{(a)})^{-1} \rho_{X \alpha v}^{(b)}$, where the inversion operates on the two last dimensions. The non-unique supercells are then discarded by removing duplicate deformation tensors, keeping the ones with the fewest atoms as priority one, the lowest norm ratio as priority two, and the smallest internal angle as priority three.
    \item Remove supercells with a maximum strain larger than $S_{\text{max}}$. This is done by minimizing the mechanical energy $E_X = \frac{1}{2} \sum_n S^{(n)}_{Xvv} C^{(n)}_{vvvv} S^{(n)}_{Xvv}$ for the common supercell $\bar{\rho}_{X \alpha v}$. Here, $C^{(n)}_{vvvv}$ is the elasticity tensor, and $S^{(n)}_{Xvv}$ the finite strain tensors between the unstrained $\rho^{(n)}_{X \alpha v}$ and common $\bar{\rho}_{X \alpha v}$ supercells given by Eq. \eqref{eq:straintensor} \cite{Schlenker:a15471}. The maximum strain $S_X^{(\text{max})}$ is then the maximum eigenvalue of $S^{(n)}_{Xvv}$.
    \item From the remaining supercells, pick the first solution with the minimum number of atoms, and transform the atoms from both layers unstrained supercells to the common supercell by applying the deformation tensor $s^{(n)}_{Xvv} = (\rho_{X \alpha v}^{(n)})^{-1} \bar{\rho}_{X \alpha v}$ to the atomic positions. Additionally, the exact twist angle can be obtained from the two deformation tensor by taking the polar decomposition of the deformation tensor, and extracting the angle from the rotation matrix obtained thereby\cite{Munn1978Polar}.
\end{itemize}

\twocolumngrid
\bibliography{bibliography} 
\end{document}


\maketitle

\section{Overview of Heterostructures}
Table 1 gives an overview of all the vdW heterobilayers included in this work. The table includes C2DB\cite{article:C2DB} unique identifiers for the two monolayers involved, their layer groups, the PBE band gaps (without spin-orbit coupling), and the amount of strain they are subjected to in the heterostructure. Additionally, the table includes the HetDB\cite{hetdb} unqiue identifier, the PBE band gap, twist angle, and number of atoms in the moiré cell of the heterostructure. Additional data on these structures is available in the HetDB online database\cite{hetdb}.
{\footnotesize
\begin{longtable}{c|cccc|cccc}
\nobreakhline
& \multicolumn{4}{c|}{Monolayer} & \multicolumn{4}{c}{Bilayer} \\
Id & Unique id & Layer group & $E_{\mathrm{gap}} \, [\mathrm{eV}]$ & Strain [\%] & Unique id & $E_{\mathrm{gap}} \, [\mathrm{eV}]$ & Twist [$^\circ$] & \#Atoms \\
\endhead
\caption{\normalsize Overview of the vdW heterobilayers considered in this work.}
\label{tab:blinfo} \\
\toprule
\nobreakhline
\nobreakhline
& \multicolumn{4}{c|}{Monolayer} & \multicolumn{4}{c}{Bilayer} \\
Id & Unique id & Layer group & $E_{\mathrm{gap}} \, [\mathrm{eV}]$ & Strain [\%] & Unique id & $E_{\mathrm{gap}} \, [\mathrm{eV}]$ & Twist [$^\circ$] & \#Atoms \\
\nobreakhline
\endfirsthead
\nobreakhline
\multicolumn{9}{r@{}}{continued \ldots}\\
\endfoot
\bottomrule
\endlastfoot
\multirow{2}{*}{1} & 2HSi-1 & p-3m1 & 2.19 & 0.02 & \multirow{2}{*}{2HSi-1\_1PbI2-1\_3} & \multirow{2}{*}{1.42} & \multirow{2}{*}{3.00} & \multirow{2}{*}{240} \\* 
 & 1PbI2-1 & p-3m1 & 2.51 & 0.02 &  &  &  & \\ 
 \hline 
\multirow{2}{*}{2} & 2HSi-1 & p-3m1 & 2.19 & 0.48 & \multirow{2}{*}{2HSi-1\_1CdBr2-1\_16} & \multirow{2}{*}{1.20} & \multirow{2}{*}{16.10} & \multirow{2}{*}{88} \\* 
 & 1CdBr2-1 & p-3m1 & 3.23 & 0.64 &  &  &  & \\ 
 \hline 
\multirow{2}{*}{3} & 2HSi-1 & p-3m1 & 2.19 & 0.44 & \multirow{2}{*}{2HSi-1\_1ClSbSe-1\_16} & \multirow{2}{*}{0.00} & \multirow{2}{*}{16.10} & \multirow{2}{*}{88} \\* 
 & 1ClSbSe-1 & p3m1 & 1.31 & 0.55 &  &  &  & \\ 
 \hline 
\multirow{2}{*}{4} & 2HSi-1 & p-3m1 & 2.19 & 0.03 & \multirow{2}{*}{2HSi-1\_1AsBrTe-1\_11} & \multirow{2}{*}{0.29} & \multirow{2}{*}{10.89} & \multirow{2}{*}{193} \\* 
 & 1AsBrTe-1 & p3m1 & 1.40 & 0.03 &  &  &  & \\
\hline 
\end{longtable}
}

\section{Tables of MADs}
Tables representing MADs and MARDs (where meaningful) are presented in this section. The tables show the differences between the different approaches to relaxing the structures. The top row presents the same values as found in the violin plots in the main text, while the following rows give further insight into two different approaches' relative consistency.
\begin{table*}[ht]
    \caption{Mean absolute deviation (MAD) and mean absolute relative deviation (MARD) of the average interlayer distance in the 336 heterobilayers relaxed with the different MLIP models and DFT with PW and LCAO basis sets. The methods are compared to each other one-by-one. In all calculations, dispersive interactions are included via the D3 correction. }
    \label{tab:z-mae}
    \centering
    \footnotesize 
    \begin{tabular}{|c|ccccccc|}
	\hline
	Structure & \multicolumn{7}{c|}{\bf{MAD} [mÅ]}\\
	relaxation method & PBE-D3(LCAO) & L-MACE-MP & M-MACE-MPA & MatterSim & DPA3 & ORB & GRACE\\
	\hline
	PBE-D3(PW)& 88.1& 133.4& 251.1& 111.0& 210.0& 110.5& 114.0 \\
	PBE-D3(LCAO)& -& 114.2& 305.7& 161.9& 210.0& 103.4& 156.5 \\
	L-MACE-MP& -& -& 324.0& 180.1& 234.3& 146.2& 192.0 \\
	M-MACE-MPA& -& -& -& 232.2& 273.6& 308.7& 217.3 \\
	MatterSim& -& -& -& -& 195.3& 177.7& 115.9 \\
	DPA3& -& -& -& -& -& 234.5& 207.5 \\
	ORB& -& -& -& -& -& -& 174.7 \\
	\hline
	 & \multicolumn{7}{c|}{\bf{MARD} [\%]}\\
	\hline
	PBE-D3(PW) & 1.38 & 2.15 & 4.36 & 1.86 & 3.51 & 1.76 & 1.95 \\
	PBE-D3(LCAO) & - & 1.85 & 5.21 & 2.67 & 3.57 & 1.66 & 2.62 \\
	L-MACE-MP & - & - & 5.49 & 2.97 & 3.95 & 2.36 & 3.24 \\
	M-MACE-MPA & - & - & - & 4.00 & 4.64 & 5.25 & 3.73 \\
	MatterSim & - & - & - & - & 3.25 & 2.91 & 1.95 \\
	DPA3 & - & - & - & - & - & 3.91 & 3.47 \\
	ORB & - & - & - & - & - & - & 2.91 \\
	\hline
\end{tabular}

\end{table*}

\begin{table*}[ht]
    \caption{Mean absolute deviations between the $z$-normalized RMSD between structures relaxed using different DFT and MLIP approaches. The $z$-normalization is performed by changing the average interlayer distance of both approaches involved to the same distance.}
    \label{tab:rmsd}
    \centering
    \footnotesize
    \begin{tabular}{|c|ccccccc|}
	\hline
	 Structure & \multicolumn{7}{c|}{\bf{MAD} [mÅ]}\\
	 relaxation method & PBE-D3(LCAO) & L-MACE-MP & M-MACE-MPA & MatterSim & DPA3 & ORB & GRACE\\
	\hline
	PBE-D3(PW) & 21.6 & 33.1 & 18.3 & 13.1 & 15.1 & 21.2 & 13.1 \\
	PBE-D3(LCAO) & - & 37.3 & 26.9 & 22.5 & 22.8 & 28.5 & 22.0 \\
	L-MACE-MP & - & - & 35.1 & 31.7 & 33.1 & 35.2 & 32.2 \\
	M-MACE-MPA & - & - & - & 16.2 & 21.1 & 26.3 & 17.2 \\
	MatterSim & - & - & - & - & 16.9 & 21.7 & 12.0 \\
	DPA3 & - & - & - & - & - & 24.4 & 17.3 \\
	ORB & - & - & - & - & - & - & 19.9 \\
	\hline
\end{tabular}

\end{table*}

\begin{table}[ht]
    \caption{Mean absolute and relative deviations between the LCAO-PBE calculated band gaps, of the structures obtained using the different approaches, MatterSim. L-MACE-MP0, and DFT with the PBE-D3 xc-functional using either a plane wave (PW) basis or LCAO basis. The relative error is always in terms of the PW relaxed structure. For the relative errors, all structures with a band gap $< 100 \, \text{meV}$ are excluded.}
    \label{tab:gap-mae}
    \centering
    \footnotesize
    \begin{tabular}{|c|ccc|}
	\hline
	Structure & \multicolumn{3}{c|}{\bf{MAD} [meV]}\\
	relaxation method & PBE-D3(LCAO) & L-MACE-MP & MatterSim\\
	\hline
	PBE-D3(PW)& 36.5& 50.5& 22.8 \\
	PBE-D3(LCAO)& -& 62.7& 42.9 \\
	L-MACE-MP& -& -& 50.0 \\
	\hline
	 & \multicolumn{3}{c|}{\bf{MARD} [\%]}\\
	\hline
	PBE-D3(PW) & 5.48 & 7.48 & 3.92 \\
	PBE-D3(LCAO) & - & 9.10 & 7.06 \\
	L-MACE-MP & - & - & 7.41 \\
	\hline
\end{tabular}

\end{table}

\begin{table}[ht]
    \caption{Mean LCAO-PBE banstructure RMSD between structures relaxed with MatterSim. L-MACE-MP0, LCAO-PBE, and PW-PBE.}
    \label{tab:bs-mae}
    \centering
    \footnotesize
    \begin{tabular}{|c|ccc|}
	\hline
	 Structure & \multicolumn{3}{c|}{\bf{MAD} [meV]}\\
	 relaxation method & PBE-D3(LCAO) & L-MACE-MP & MatterSim\\
	\hline
	PBE-D3(PW) & 40.8 & 59.8 & 34.5 \\
	PBE-D3(LCAO) & - & 61.7 & 46.2 \\
	L-MACE-MP & - & - & 57.3 \\
	\hline
\end{tabular}

\end{table}
\newpage
\printbibliography